# Freezing of a Liquid Marble

*Ali Hashmi, Adam Strauss, and Jie Xu\**

Mechanical Engineering Department, Washington State University, Vancouver

ABSTRACT

In this study, we present for the first time the observations of a freezing liquid marble. In the experiment, liquid marbles are gently placed on the cold side of a Thermo-Electric Cooler (TEC) and the morphological changes are recorded and characterized thereafter. These liquid marbles are noticed to undergo a shape transition from a spherical to a flying-saucer shaped morphology. The freezing dynamics of liquid marbles is observed to be very different from that of a freezing water droplet on a superhydrophobic surface. For example, the pointy tip appearing on a frozen water drop could not be observed for a frozen liquid marble. In the end, we highlight a possible explanation for the observed morphology.

INTRODUCTION

Liquid marbles is a relatively new area of research that is rapidly finding applications in many fields, such as material science and chemical engineering.[1] Liquid marbles are a species comprising of a liquid-filled core enveloped by micro or nanometrically scaled particles. These particles can be hydrophobic or hydrophilic, such as graphite,[2] carbon black,[3] OTFE/PTFE,[4] lycopodium,[5] PFA-C$_8$,[6] silica aerogels,[7] silica nanoparticles,[8] cellulose powder,[9] and many others.[10] Liquid marbles have particularly shown promise as chemical sensors,[11] micropumps,[12] and microreactors.[13] Recently, various novel coatings have also been employed for liquid marbles to make them responsive to pH,[14] electric field[15] and magnetic field,[16] and it is hoped that new materials and coatings for liquid marbles will enable us to achieve even exciting applications.[1] However, liquid marble research is still in its infancy, and many fundamental characteristics of liquid marbles are yet to be explored,[17] such as the phase change processes of a liquid marble. The evaporation of a liquid marble has been previously studied by Erbil group [18-19] and our group[20]. It has been found that liquid marbles are thermodynamically analogous to Leidenfrost droplets since the coating surrounding the liquid marbles acts as a barricade preventing heat transfer from the hot surface to the liquid core. The process of condensation into liquid marbles has also been reported recently.[21] In this letter, we report for the first time observation of the freezing of liquid marbles. Unlike frozen liquid droplets, liquid marbles acquire a flying-saucer shaped morphology upon freezing. We also present a possible explanation for the morphology of a frozen liquid marble.

EXPERIMENTAL

The experimental setup used is simple and comprises of a Thermo-Electric Cooler (TEC), also called Peltier cooler, (SuntekStore, 100W, 12V) connected to a DC voltage supply (TDK-Lambda, GEN 150-5) for cooling purpose. A silicon wafer (Universitywafer.com, P/B(100), contact angle measured to be about $30^o$) is attached to the cold side of the TEC using a thermal conductive paste (Omega, OT-201), which ensures rapid heat transfer between the TEC and the silicon wafer. Likewise, a layer of thermal conductive paste is applied between the hot side of the TEC and a large aluminum block (3x3x5 $cm^3$) surrounded by a heat sink (ice-water mixture) ensuring that the stray heat generated at the hot side of TEC is readily conducted away. A high speed camera (PixeLink, PL-B771U) is used to monitor and record the shape transition during the freezing process. In our experiments, liquid marbles are created using lycopodium powder (Fisher Scientific). Optical microscopy (Figure. 1) reveals that the lycopodium particles are nearly spherical in nature with an average diameter of 28 ± 5 μm.

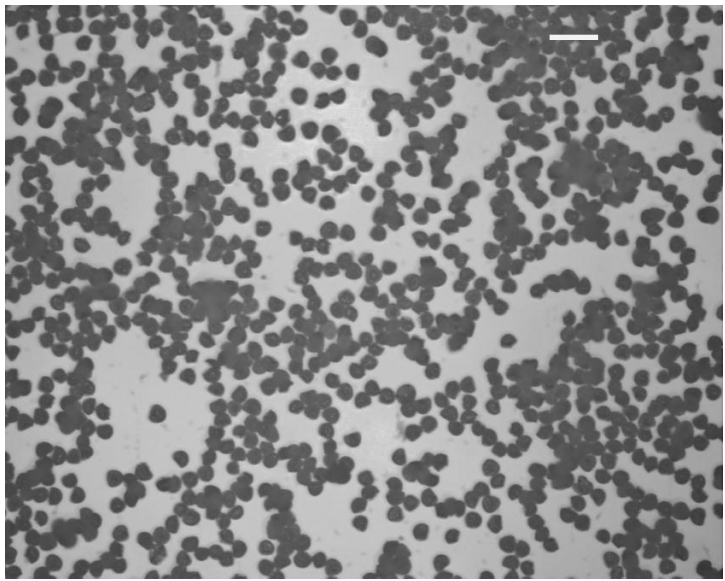

Figure 1. An optical image of lycopodium particles. The top-right scale bar represents 100 μm.

In order to prepare the liquid marble, some lycopodium powder is first added onto a glass slide and thereafter a de-ionized (DI) water droplet is dripped from a syringe onto the pile of lycopodium particles, then rolled over the powder back and forth to be thoroughly coated, creating a stabilized liquid marble. The usage of syringe enables a controlled release of DI water resulting in the generation of variously sized liquid marbles. A thermocouple (Omega, K type) is used to monitor the surface temperature of the silicon wafer, which was kept at approximately –8 °C throughout the experiments.

Before liquid marbles are prepared and placed on top of the silicon wafer attached to the cold side of the TEC, appropriate temperature conditions need to be ensured. For this, the current flowing through the TEC is adjusted and several seconds is allowed until a desired steady state temperature reading is attained. Subsequently, a liquid marble is gently placed on top of the cold silicon wafer and video recordings are made from the instance of contact until the formation of a frozen liquid marble or an ice marble hereafter.

OBSERVATIONS AND DISCUSSION

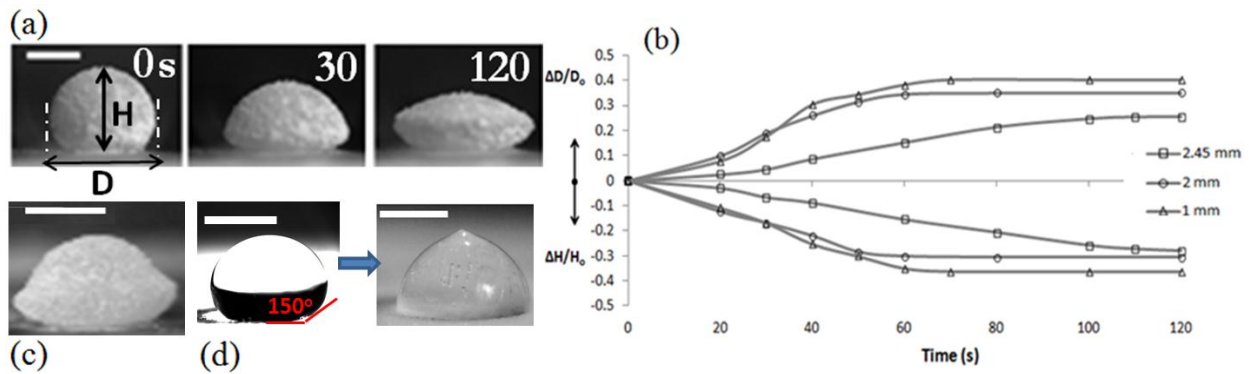

Figure 2. (a) Liquid marble freezing toward a flying saucer shaped ice marble. (b) The relative change in height ($\Delta H/H_o$) and diameter ($\Delta D/D_o$) of different sized liquid marbles with time. The graph delineates that a smaller-sized liquid marble freezes faster compared to a larger marble. (c) A bell-shaped frozen liquid marble. (d) A water droplet on a superhydrophobic surface – before and after freezing (note that a pointy tip can be seen on top of the frozen droplet – a feature absent in the frozen liquid marbles). The scale bars represent 1 mm.

During initial moments of contact (t = 0 seconds) smaller liquid marbles are noticed to almost maintain a spherical shape (or a puddle shape for larger marbles) on the silicon wafer (Figure 2a, left). However, during the cooling process we note that liquid marbles experience a deformation in their shape and their spherical morphologies undergo a gradual transition to a flying saucer shaped pattern, which we have found to be very repeatable. A decrease in the height of the liquid marbles (ΔH) is accompanied by a bulging of the sides (ΔD). From Figure 2a, middle, it is apparent that bulging takes place in the region closest to the contact area (solid-solid interface). The liquid marble by this time loses its sphericity and attains a dome shape. Upon further cooling, the height of the liquid marble is further lowered and the sides extend outwards forming a shape that resembles an ellipsoid or a flying saucer (see Figure 2a, right). During the cooling process, the freezing front is seen to initiate from the bottom and move gradually toward the top of the liquid marble. The time required for the formation of ice marbles is a few minutes for the typical substrate temperature used in our experiments depending upon the mass of the liquid marble. For larger liquid marbles, a similar sideways expansion (ΔD) and height-shrinkage phenomenon (ΔH) is observed and reported in Figure 2b (D and H are defined in Figure 2a). Very occasionally, we observed bell shaped liquid marbles, as shown in Figure 2c. For all the tested cases, the top of the frozen ice marbles is observed to be smooth, without any signs of pointy structures, cusps or a singularity. To make a direct comparison to the freezing of water droplets, a superhydrophobic surface is conveniently fabricated using Teflon sheet and a sandpaper.[22] The contact angle of water on the superhydrophobic surface is measured to be around 150º (see Figure 2d). In the experiment, a water droplet cooled on the superhydrophobic surface is noticed to expand upwards during freezing, forming a small singularity at the top (Figure 2d); no sideways expansion is observed, although the contact angle seems to have decreased after freezing. All these observations are very repeatable and are consistent with previous reports.[23-24] It is worth noting that freezing on a superhydrophobic surface has recently drawn extensive attention primarily due to the anti-icing effects of the surface.[25]

In case of liquid marbles, the observed deformation may very possibly be explained on the basis of Marangoni convection and preferred nucleation. Thermodynamically, because only the bottom layer of particles are directly cooled by the cold substrate, heterogeneous nucleation is expected to start from the inside of the marble near the bottom layer of particles where the liquid molecules have a lower energy barrier to cross over in order to initiate nucleation; this energy barrier for nucleation, which is best known as the Gibbs free energy, $\Delta G$, can be estimated as:[26]

$$\Delta G = \frac{16\pi \gamma_{sl}^3 T_{slf}^2}{3 h_{sl}^2 (T_{slf} - T_{interface})^2} s(\theta_c) \quad (1)$$

where

$$s(\theta_c) = \frac{1}{2} + \frac{3}{4}\cos\theta_c - \frac{1}{4}(\cos\theta_c)^3 \quad (2)$$

In the above equations, $\gamma_{sl}$ is the particle(solid)-liquid interfacial energy, $T_{slf}$ is the temperature of the freezing or solidification front inside the marble, $h_{sl}$ is the latent heat of vaporization, $T_{interface}$ is temperature of the interface between the liquid core and the bottom layer of particles, and the factor $s(\theta_c)$ accounts for the effects of heterogeneous nucleation where $\theta_c$ is the contact angle of ice on particle surface in water ambient. The transient $T_{interface}$ is approximated by solving heat conduction equations in both particle domain and the liquid core domain, and is given by the following:

$$T_{interface}(t) = T_{marble} + (T_{substrate} - T_{marble}) \frac{erfc\left(\frac{h}{2\sqrt{\alpha_{particle}t}}\right)}{erfc\left(\frac{h}{2\sqrt{\alpha_{water}t}}\right)} \quad (3)$$

where $T_{marble}$ is the average temperature of the liquid marble at t = 0 sec, $T_{substrate}$ is the temperature of the substrate, $h$ is thickness of the marble coating in contact with the substrate (assumed to be of uniform thickness), $\alpha_{particle}$ and $\alpha_{water}$ are the thermal diffusivities of the particles and water respectively, with the sparse coating (liquid marble shell) assumed to have the same thermal diffusivity as the bulk particle material. The above equations describe the freezing dynamics which includes nucleation and heat transfer from marble to the substrate across a layer of particles. However, precise evaluation of these equations seems to be difficult at this stage. This is mainly due to the experimental difficulty in determining multiphase interfacial temperatures, which is proven to be a challenging task.[27] After the nucleation starts at the bottom of the marble, it will grow inside the marble (see Figure 3). Owing to the higher temperatures of the coating particles (except those particles at the bottom of the marble that are cooled directly on the substrate), the energy barrier is significant at the particle-liquid interface away from the substrate and thus, this may inhibit nucleation process on the marble surface as can be explained from equation (4).[28]

$$J(t) = K \exp(-\frac{\Delta G}{kT_{interface}}) \quad (4)$$

where, $J$ represents the rate of nucleation, $k$ is the Boltzmann constant and $K$ is the kinetic coefficient. Keeping this in view, a liquid zone would tend to exist between the freezing front and the coating during the freezing process. In short, we postulate that the particle coating on the liquid marble acts as an icing-inhibitor by providing thermal energy to the adjacent liquid molecules, which implies that nucleation preferentially takes place on the inside of the marble near the substrate. A surface tension driven flow owing to temperature gradients, commonly known as Marangoni convection,[29] exists if the dimensionless Marangoni number is above the critical threshold of 80.[30] Experiments in our case with the typical marble geometric characteristics approximates to a Marangoni number on the order of 10, 000 which corresponds to a strong Marangoni convection. Note that even though the surface tension of the marble is lower than pure DI water alone, [31] it should not have any significant effect on the Marangoni flow since the estimated Marangoni numbers are far above the threshold. This Marangoni effect will induce liquid flow into the gap between the coating and the freezing front, as sketched in Figure 3. This may explain why bulging is observed only at the sides of the marble accompanied with a decrease in the marble's height. Moreover, the direction of Marangoni flow is from the marble top to marble sides around the inner ice core; therefore, this flow helps transport liquid away from the top of the marble, and in turn contributes to the absence of pointy peaks in ice-marbles. Experiments are also repeated by placing the liquid marbles on top of the TEC and then inverting it upside-down i.e. against the direction of gravity. Similar experimental observations of freezing patterns point to the fact that gravity has a negligible effect on the freezing process.

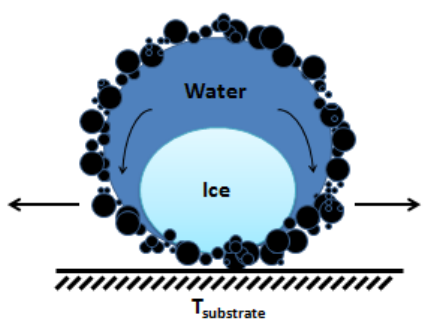

Figure 3. The sketch shows the cross-section of the liquid marble with the proposed gap between the freezing front and the coating due to preferential nucleation. The arrow indicates the flow of water to the gap due to Marangoni effects.

Unlike liquid marbles, there are no preferential nucleation sites and no gap exists between the water-air interface and the freezing front in case of freezing water droplets, which might be the reason for the absence of a sideways convection. As a consequence, water in the vicinity is only able to see the freezing front and freezes which results in an upward expansion of the water droplet that finally converges to a pointy peak. The speed at which the freezing front propagates in case of a freezing water droplet is observed to be relatively higher than that in liquid marbles.

In order to investigate the effects of the absence of coating from the base of a liquid marble on the freezing dynamics, we carried out a second set of experiments. Water droplet was added on top of the silicon wafer and some lycopodium powder was then coated on top of the water droplet, which was then allowed to freeze. During the freezing process we observe a fast moving freezing front travelling from the bottom (water-silicon interface) toward the top; a rapid upward and a slight sideways expansion is observed during the initial period; however, in the last phase of freezing a crater appears on top of the "half-marble" (Figure 4 left), as opposed to a pointy tip or a cusp in case of a freezing water droplet on the same surface (Figure 4, right).

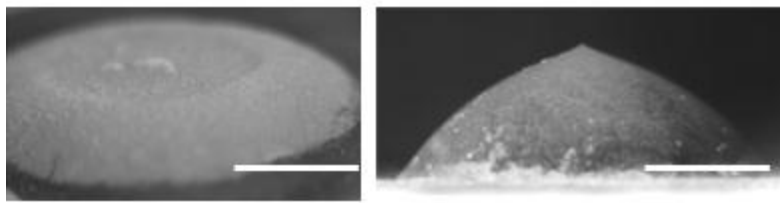

Figure 4: (left) Formation of a crater upon freezing of a half-marble: sessile water droplet coated with microparticles. (right) a pointy peak forms when a water droplet freezes on the same surface. The scale bars represent 1 mm.

The slight sideways expansion and the formation of a crater at the completion of the freezing process of a "half-marble" may probably be due to the same effect as previously described: a gap between the water-particle interface and the freezing front receives water flow from Marangoni effects. We again repeated the experiments with the system inverted, and observed the same freezing pattern indicating that the morphological changes are not influenced by gravity.

CONCLUSIONS

In summary, we performed experiments to monitor the dynamics of a freezing liquid marble. We note that the morphology of a liquid marble with water-filled core is very different from a water droplet freezing on a supercooled substrate. The flying saucer shaped ice marble is formed possibly due to the icing-inhibition characteristics of the hydrophobic microparticles enveloping the liquid core, and the Marangoni convection causing the liquid to flow and seep down sideways around the marble. Likewise, we carried out experiments and observed that a water droplet residing on a silicon substrate coated with microparticles results in the formation of a crater upon freezing, as opposed to pointy protrusion formed on a frozen water droplet.

AUTHOR INFORMATION

**Corresponding Author**

*Tel: 1-360-546-9144. Fax: 1-360-546-9438. E-mail: jie.xu@wsu.edu.

Table of Contents only:

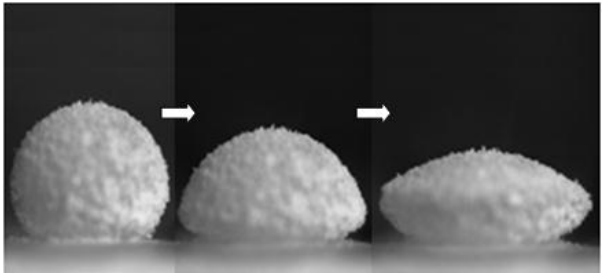